\documentclass[aps,pra,twocolumn]{revtex4}

\usepackage{graphicx}
\usepackage{amssymb}
\usepackage{amsfonts}
\usepackage{epstopdf}

\DeclareGraphicsExtensions{.eps}

\begin{document}

\title{Atomic-ensemble-based quantum memory for sideband modulations}

\author{J. Ortalo, J. Cviklinski, P. Lombardi, J. Laurat, A. Bramati, M. Pinard, E. Giacobino}

\address{Laboratoire Kastler Brossel, Universit\'{e} Pierre et Marie Curie, Ecole Normale Sup\'{e}rieure, CNRS,
Case 74, 4 place Jussieu, 75252 Paris Cedex 05, France. \\E-mail:
elg$@$spectro.jussieu.fr}

\begin{abstract}
Interaction of a control and a signal field with an ensemble of
three-level atoms allows direct mapping of the quantum state of
the signal field into long lived coherences of an atomic ground
state. For a vapor of cesium atoms, using Electromagnetically
Induced Transparency (EIT) and Zeeman coherences, we compare the
case where a tunable single-sideband is stored independently of
the other one to the case where the two symmetrical sidebands are
stored using the same transparency window. We study the conditions
in which simultaneous storage of two non-commuting variables
carried by light and subsequent read-out is possible. We show that
excess noise associated with spontaneous emission and spin
relaxation is small, and we evaluate the quantum performance of
our memory by measuring the signal transfer coefficient T and the
conditional variance V and using the T-V criterion as a state independent benchmark.

\end{abstract}

\maketitle

\section{Introduction}
Quantum information technology is a fast developing field,
which aims to exploit  new ways for
information processing and communication with no analogue in
classical information science. This paradigm leads in particular towards absolutely
secure communications and computing powers beyond the
capacities of any classical computer. Quantum
communications, in particular scalable networks, as well as quantum computation rely critically on memory
registers and considerable interests are currently dedicated to this quest  \cite{zoller,kimble}.

Developing registers for quantum variables requires however
completely different concepts than for classical data. In such
systems, photonic quantum data must be linearly and reversibly
converted into long-lived matter states. Atomic ensembles are good
candidates for such memory  since quantum states of light can be
stored in long-lived atomic spin states by means of light-matter
interaction and retrieved on demand. Starting around the year
2000, several protocols have been proposed for light-matter
interfacing.  Different configurations have been studied,
including Raman-type configuration \cite{Polzik1,DLCZ,Dantan1},
resonant electromagnetically induced transparency (EIT)  scheme
\cite{Lukin1,Lukin2,Dantan2} or quantum non demolition (QND)
interaction using a X-type transition \cite{Polzik2}. Rephasing
photon echo or spin echo interactions for storage and read-out of
the quantum signal have also been studied
\cite{Moiseev,Kraus,Sangouard}.

Considerable experimental works have been performed in the single photon regime,
including probabilistic protocols in Raman configuration \cite{james, julien, james2,Pan,Vuletic} or storage and read-out of single photons using
the EIT scheme \cite{Lukin3,Kuzmich}. Recent papers present notable advances
with reversible mapping of single-photon entanglement \cite{Choi} and enhanced
memory time \cite{Zhao}. For future quantum communications,
continuous variables may offer improved data transmission rates
\cite{Grangier}, and, in this paper, we will focus on this regime. Storage of non-commuting quantum
variables of a light pulse has been demonstrated in 2004
\cite{Polzik3}, however without the possibility of read-out. Significantly, very
recent results demonstrated storage and retrieval of a squeezed
light pulse, or of faint coherent state with quantum performances
\cite{Kozuma,Lvovsky,Cviklinski}.

Presently, while advances have been achieved in the direction of a
deterministic quantum memory, it is interesting to investigate a
variety of different systems allowing more flexibility. In this
paper, we compare the potential of the EIT scheme when quantum
sidebands are stored independently, to the case where two
sidebands are stored in the same EIT window. In particular, we
show that when single sidebands are stored on an adjustable
frequency range, using the Zeeman coherence of the atoms, the
optimal response of the medium for storage can be adapted to the
frequency to be stored by changing the magnetic field, while
keeping a narrow EIT window. If symmetrical sidebands are stored
in separate atomic ensembles, this method should allow the storage
of a variety of quantum signals. This opens the way to quantum
storage and retrieval of  quantum fluctuations with adjustable
frequency.

\begin{figure*}[t!]
\centering
\includegraphics[width=0.285\textwidth]{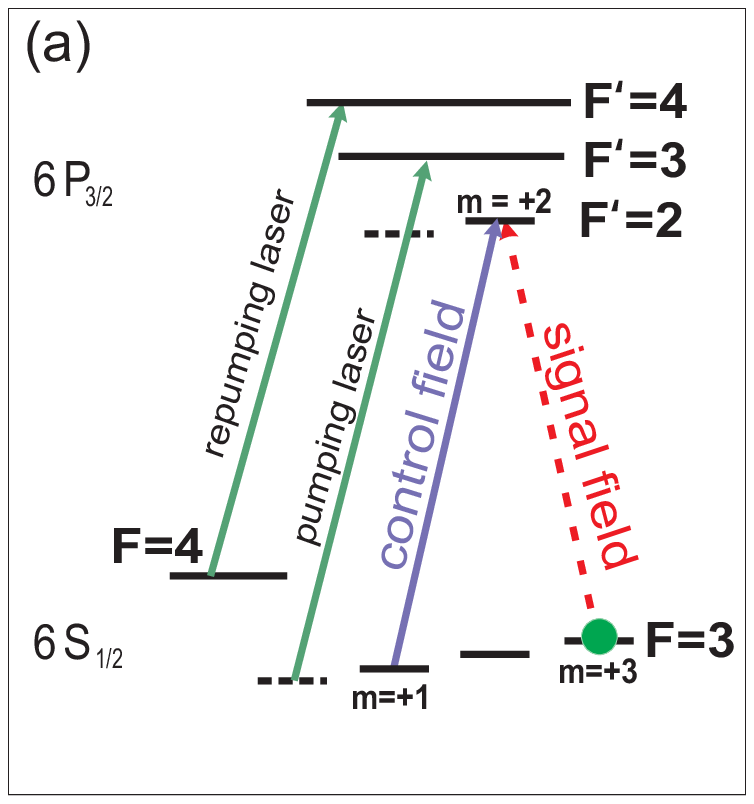}
\includegraphics[width=0.5\textwidth]{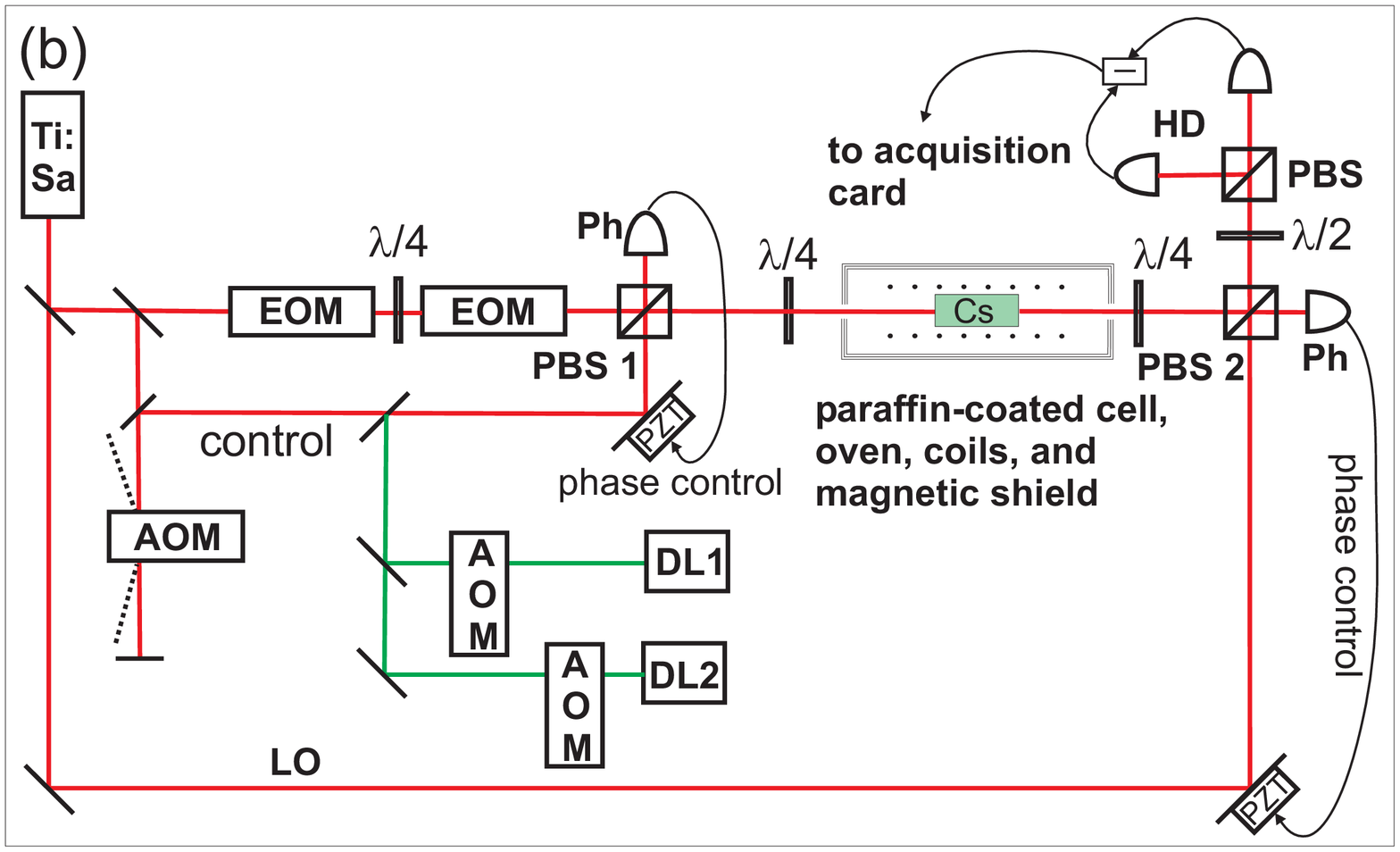}
\caption{(a) Cesium atomic levels and transitions involved in the
experiment. (b) Schematic diagram of the experimental setup. The
Ti-sapphire laser generates a beam which is used as the control
beam; a part of it is split off and frequency shifted by  two
electro-optic modulators (E.O.M.) to generate the signal beam .
The two beams are recombined on a polarizing beamsplitter (PBS1)
and sent into the cell together with the pumping and repumping
beams produced by diode lasers DL1 and DL2,  the frequencies of
which are adjusted with acousto-optic modulators (A.O.M.). The
beam going out of the cell is mixed with a local oscillator (with
the same frequency as the control beam) on polarizing beamsplitter
PBS2 and sent to a homodyne detection (H.D.)} \label{setup}
\end{figure*}

\section{EIT-based quantum memory}

We consider a large ensemble of N three-level atoms in a $\Lambda$
configuration, with two ground states and one excited state. The
storage protocol relies on two light fields, the control field and
the signal field, that interact with the atoms, with frequencies
close to resonance with the two atomic transitions. For optimal
efficiency, the two-photon resonance must be fulfilled. The control
field is a strong, classical field that makes the medium transparent
by way of EIT for the signal field to be stored \cite{Harris}. The
signal field is a very weak coherent field. If the control field is
strong enough, the atomic medium becomes transparent for the signal
field, and the refractive index acquires a very high slope as a
function of frequency in the vicinity of the one-photon resonance.
Thus the group velocity for the signal field is strongly reduced
\cite{Hau1,Hau2,Phillips}.

When the signal pulse is entirely inside the atomic medium and
after a write time, which is on the order of the characteristic
interaction time between atoms and fields, the control field can
be switched off. The two quadratures of the signal are then stored
in two components of the ground state coherence. For read-out, the
control field is turned on again. The medium emits a weak pulse,
similar to the original signal pulse, that goes out of the medium
together with the control field.

Models developed in Refs \cite{Lukin1,Lukin2,Dantan2} predict that
the classical mean values as well as the quantum variables of the
field can be stored with a high efficiency as collective spin
variables, and that the quantum variables can be retrieved in the
outgoing pulse with a very good efficiency
\cite{Dantan3,Sorensen}. Taking into account all the noise
sources, including the atomic noise generated by spontaneous
emission and spin relaxation, a significant result of theoretical
models \cite{Dantan2,Dantan3,Sorensen} is the absence of excess
noise in the process of storage and retrieval, provided the
optical depth of the medium is large enough.

\begin{figure*}[htpb!]
\begin{center}
\includegraphics[width=0.8\textwidth]{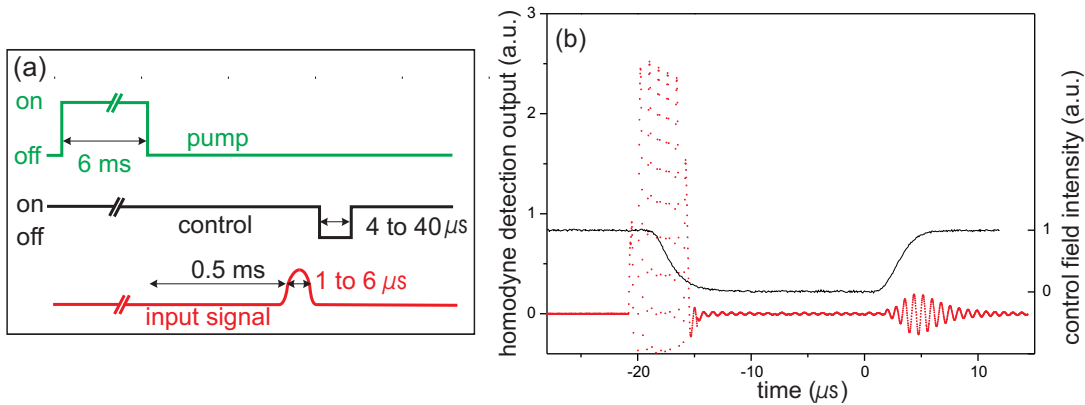}
\end{center}
\caption{(a) Detail of the experimental
sequence. (b) Temporal profile of the control field during the writing,
storage and reading stage (solid line), together with the homodyne
detection output current (dots) for a macroscopic 4~$\mu$W input
pulse, in arbitrary units.} \label{sequence}
\end{figure*}

\section{Experimental scheme}

The level configuration is shown in Fig. \ref{setup}(a). It
involves the transition $6S_{1/2}, F=3$ to $6P_{3/2}, F'=2$ of the
Cesium D$_{2}$ line. More specifically, the  control beam is
$\sigma^{+}$ polarized and resonant with the $m_{F}=1$ to
$m_{F'}=2$ transition while the signal field is $\sigma^{-}$
polarized and resonant with the $m_{F}=3$ to $m_{F'}=2$ transition
(Fig. \ref{setup}(a)). In order to fulfill the two-photon
resonance condition, the detuning $\Omega$ between the control and
signal beams is set to be equal to the Zeeman shift $2\Omega_{L}$
between the $m_{F}=1$ and $m_{F}=3$ sublevels of the ground state

\begin{equation}
\Omega = 2\Omega_{L} = 2\mu_{B}gH \label{Zeeman}
\end{equation}
where $g$ is the Land\'{e} factor, $\mu_{B}$ the Bohr magneton and
$H$ is the magnetic field. By tuning the magnetic field, one can
optimize the memory response for a given frequency, allowing a
widely tunable frequency range for the signal to be stored.

The experimental set-up is sketched on Fig. \ref{setup}(b). The
Cesium vapor is contained in a 3 cm long cell with a paraffin
coating that suppresses ground state decoherence caused by
collisions with the walls. The cell is heated to temperatures
ranging from 30$^{o}$C to 40$^{o}$C, yielding optical depths from
6 to 18 on the signal transition. It is placed in a longitudinal
magnetic field adjustable between 0 and 2~Gauss produced by
symmetrical sets of coils and in a magnetic shield made of three
layers of $\mu$metal. Residual magnetic fields are smaller than
0.2~mG, and the homogeneity of the applied magnetic field over the
cell volume, measured by magneto-optical resonance, is better than
1:700. The atoms are optically pumped from the F=4 to the F=3
ground state and into the $m_{F}=3$ sublevel of the F=3 ground
state with diode lasers, as shown in Fig. \ref{setup}(b). The
control and signal beams are produced by a single-mode,
Ti-Sapphire laser, with a linewidth of 100 kHz, stabilized on a
saturated transition absorption. The control beam fluctuations are
at the shot-noise limit for frequencies above 1 MHz.

The signal beam is produced by splitting off a part of the control
beam and generating a single sideband shifted from the initial
frequency by using a set of two electro-optic modulators separated
by wave plates and polarizers, as described in \cite{Cusack}. The
second sideband is suppressed by 20~dB. The single sideband is a
very weak coherent field, with a power on the order of a nanowatt,
with an adjustable frequency detuning $\Omega$ and with a
polarization perpendicular to that of the carrier. The carrier is
filtered out by reflection on a polarizing beam splitter (PBS1 in
Fig. \ref{setup}(b)), the reflected beam being used to lock the
signal to control relative phase. The control field from the
Ti-Sapphire laser, with a power of 10 to 140~mW, can be turned on
and off with an acousto-optic modulator. It has a vertical
polarization, and it is mixed with the horizontally polarized weak
signal beam on the same polarizing beamsplitter. The beam is sent
into the cell after passing through a quarter-wave plate, which
produces the $\sigma^{+}$ and $\sigma^{-}$ polarizations.

\section{Detection}

The light going out of the cell is mixed with a local oscillator and
analyzed using a homodyne detection, after eliminating the control
beam by means of a polarizing beamsplitter (PBS2 in Fig.
\ref{setup}(b)). The local oscillator is obtained from the same
Ti-Sapphire laser as the control and signal beams and has the same
frequency as the control beam. Its phase is locked to the one of the
control beam.

The photocurrent difference obtained after the homodyne detection
yields the amplitude modulation operator $\hat{X}(\Omega)$:

\begin{equation}
\hat{i}=\hat{X}(\Omega)=(\hat{X}_{\Omega}+\hat{X}_{-\Omega})\cos\Omega
t+ (\hat{Y}_{\Omega}-\hat{Y}_{-\Omega})\sin\Omega t
\label{photocur}
\end{equation}
which is expressed as a combination of the quadrature operators of
the two sidebands, $\hat{X}_{\pm\Omega}$ and
$\hat{Y}_{\pm\Omega}$. Since the two operators
$\hat{X}_{\Omega}+\hat{X}_{-\Omega}$ and
$\hat{Y}_{\Omega}-\hat{Y}_{-\Omega}$ commute, one can measure them
at the same time on the sine and cosine components. However, the
modulated signal field has only one sideband, so the components
$\hat{X}_{-\Omega}$ and $\hat{Y}_{-\Omega}$ at $-\Omega$ are
empty, corresponding to one unit of shot noise in the opposite
sideband, which is an intrinsic feature of homodyne detection in
this case.

In the experimental timing shown in Fig. \ref{sequence}(a), the
atoms are first pumped into the $m_{F}=3$ sublevel of the
$6S_{1/2}, F=3$ level, using a $\sigma_{+}$ pumping diode laser,
with a power of 0.2~mW, in the presence of an additional diode
laser repumping the atoms from the $6S_{1/2}, F=4$ level to the
$6S_{1/2}, F=3$ level, with a power of 2~mW. The pumping stage
lasts for 6 ms, after which more than 90$\%$ of the atoms are
pumped into the $m_{F}=3$ sublevel, as shown in Fig. \ref{setup}
(a). After a dark period of 0.5~ms for the pumping and repumping
laser diodes, the signal pulse is sent into the cell for the
writing procedure, the control field being on. The signal pulse
can last for 1 to 6~$\mu$s. The control and signal fields are then
switched off for 4 to 40~$\mu$s, and the control field alone is
turned on again. Fig. \ref{sequence}(b) shows the beat signal of
the light going out of the cell with the local oscillator. The
first pulse corresponds to the signal field transmitted during the
writing period, the second one to the field read out from the
signal stored by the atoms.

The photocurrent difference from the homodyne detection is recorded
at a rate of 50 Megasamples per second with a 14 bits acquisition
card (National Instruments NI 5122). A Fourier transform is
performed digitally by multiplying the signal with a sine or a
cosine function of angular frequency $\Omega$ and integrating over a
time $t_{m}=n 2 \pi /\Omega$, with $n$=2 to 4. This yields sets of
measured values of the quadrature operators $\hat{X}_{\Omega}$ and
$\hat{Y}_{\Omega}$ of the outgoing field. Averaging over 2000
realizations of the experiment gives direct access to the quantum
mean values $<\hat{X}_{\Omega}>$ and $<\hat{Y}_{\Omega}>$ and
variances
$<(\Delta\hat{X}_{\Omega})^2>=<(\hat{X}_{\Omega})^{2}>-(<\hat{X}_{\Omega}>)^{2}$
and
$<(\Delta\hat{Y}_{\Omega})^2>=<(\hat{Y}_{\Omega})^{2}>-(<\hat{Y}_{\Omega}>)^{2}$
of the signal field quadratures.

\begin{figure}[htpb!]
\centering
\includegraphics[width=0.8\columnwidth]{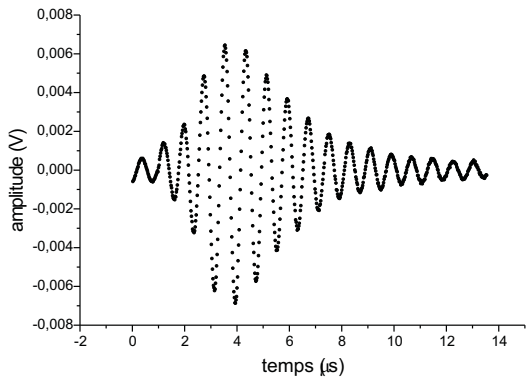}
\caption{Typical temporal profile used for demodulation in order
to enhance the  measurement efficiency.} \label{LO}
\end{figure}

\begin{figure}[b!]
\centering
\includegraphics[width=0.97\columnwidth]{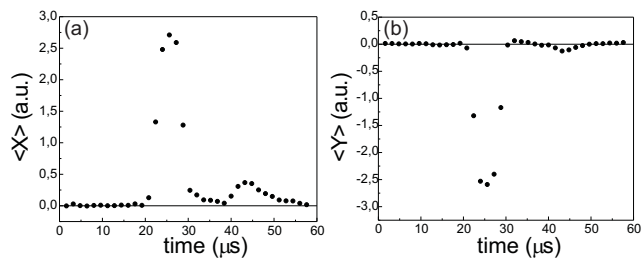}
\caption{Time-dependent mean values of the amplitude (a) and phase
(b) quadratures, measured from a typical 2000-sequence run (see text
for details). T=36$^\mathrm{o}$C, control field power = 9~mW, input
pulse duration : 5~$\mu$s, input intensity : 0.1~nW.} \label{signal}
\end{figure}

In order to improve the efficiency of the measurement of the
output signal, the temporal profile of the local oscillator must
be adapted to the one of the signal \cite{Dantan3}. In an
equivalent way, one can adapt the profile of the sine or cosine
function used during demodulation. Figure \ref{LO} represents the
optimal shape of the function used during the Fourier transform
yielding the values of $\hat{X}$ and $\hat{Y}$. Using this profile
instead of a usual sine function with a constant amplitude
improves the measurement efficiency by a factor 3, because it
avoids measuring the vacuum field outside the relevant time
interval of the pulse. This improved profile is created by
recording the signal going out of the memory in the case of a
macroscopic signal (with a power of a few microwatts).

\section{Experimental results}

Typical traces are shown in Fig. \ref{signal}. The first peak
corresponds to the transmitted part of signal field, the second part
to the retrieved signal, for the in-phase ($X$) quadrature (curve
(a)) and for the out-of-phase ($Y$) quadrature (curve (b)).

\begin{figure}[htpb!]
\centering
\includegraphics[width=0.8\columnwidth]{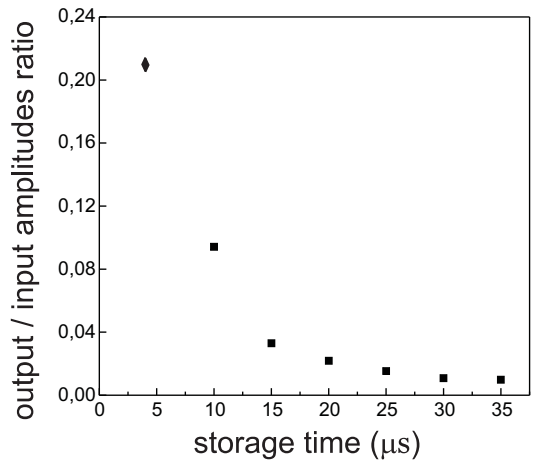}
\caption{Ratio of the amplitudes of the input and output states, as a
function of the storage time. T=40$^\mathrm{o}$C. The points
indicated by $\blacksquare$ $(\blacklozenge)$ correspond
to a signal pulse duration of 6.4 $\mu \mathrm{s}$ (1.6 $\mu
\mathrm{s}$) with a control field power of 10~mW (140 mW).}
\label{temps_mem}
\end{figure}

As far as the mean values are concerned, Fig. \ref{signal} shows
that the experiment allows to store the two quadratures of a
signal in the atomic ensemble and then to retrieve them. The
retrieved signal is about 10$\%$ of the transmitted signal in
amplitude. The
storage efficiency as a function of the storage time is shown in
Fig. \ref{temps_mem}. The storage efficiency decreases rapidly
with the storage time, with a time constant $\tau_{m} \sim 10
\,\mu$s, due to spin relaxation in the ground state, in particular
because of stray magnetic fields and collisions. An efficiency of
21$\%$ is measured for a short storage time with a strong control
field.

\begin{figure}[htpb!]
\centering
\includegraphics[width=0.97\columnwidth]{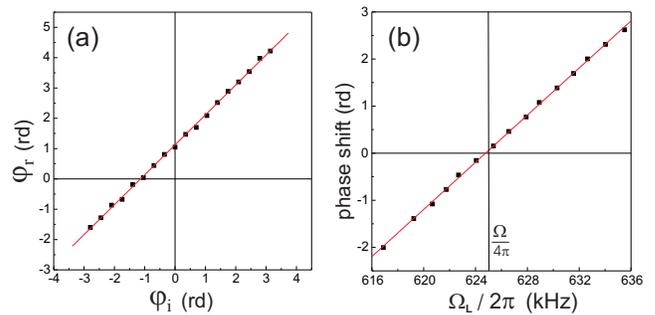}
\caption{Dependence of the retrieved pulse phase, as a function of
the input pulse phase with a constant Larmor frequency (a), and as a
function of the Larmor frequency (b) (i.e : as a function of the
two-photon detuning), with a fixed phase for the input phase. The
storage time is 20~$\mu$s, and $\Omega / 2\pi=$1.25 MHz.}
\label{phase}
\end{figure}

In order to check the phase coherence of the process, we have
performed a detailed study of the phase of the retrieved signal
\cite{coherence}. When the two photon transition resonance with the
control field and  the signal field is fulfilled,
$\delta=2\Omega_{L}-\Omega=0$, the atomic coherence evolves during
the storage time with a frequency which is equal to $\Omega$. When
the control field is sent again into the atomic ensemble for
read-out (with a phase that is still locked to the phase of the
local oscillator), the outgoing field is expected to be emitted with
the same phase relative to the control field as the one of the
signal field. If the two-photon transition is slightly off
resonance, the atomic coherence accumulates a phase difference
during the storage time. This causes a phase shift of the retrieved
signal as compared to the control field.

Figure \ref{phase}(a) shows the measured dependence of the phase
$\varphi_{r}$ of the retrieved pulse on the phase $\varphi_{i}$ of
the initial pulse. Phase $\varphi_{r}$ shows a linear dependence
on $\varphi_{i}$ with a unit slope, confirming the coherence of
the process, while the non-zero value of $\varphi_{r}$ for
$\varphi_{i}=0$ corresponds to a small two-photon detuning in the
storage process. We show in figure \ref{phase}(b) the measured
phase shift of the retrieved signal as a function of $\delta$ by
varying the magnetic field for a fixed storage time. The phase
shift shows a linear dependence on the Larmor frequency, with a
slope of 0.25 rd/kHz which is in very good agreement with the
predicted dependence, given by
$\varphi_{r}=(2\Omega_{L}-\Omega)\tau $, where $\tau$= 20~$\mu $s
is the storage time. These measurements show the full phase
coherence of the process.

\begin{figure}[htpb!]
\centering
\includegraphics[width=0.9\columnwidth]{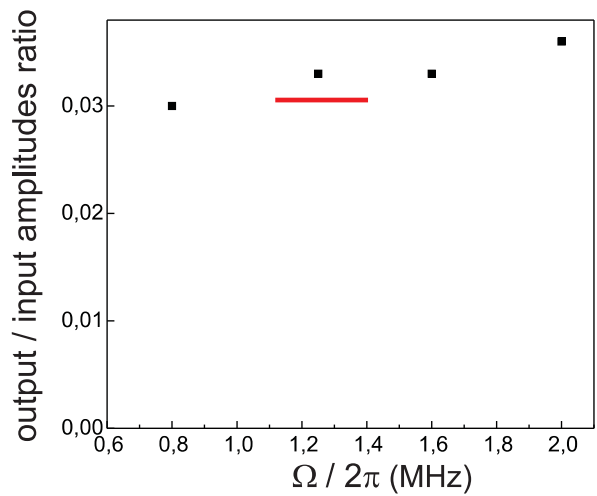}
\caption{Ratio of the amplitudes of the input and output states,
as a function of the modulation frequency $\Omega / 2 \pi$. The
bar indicates the spectral width of the input pulse.} \label{st}
\end{figure}

\begin{figure*}[ht]
\centering
\includegraphics[width=0.70\textwidth]{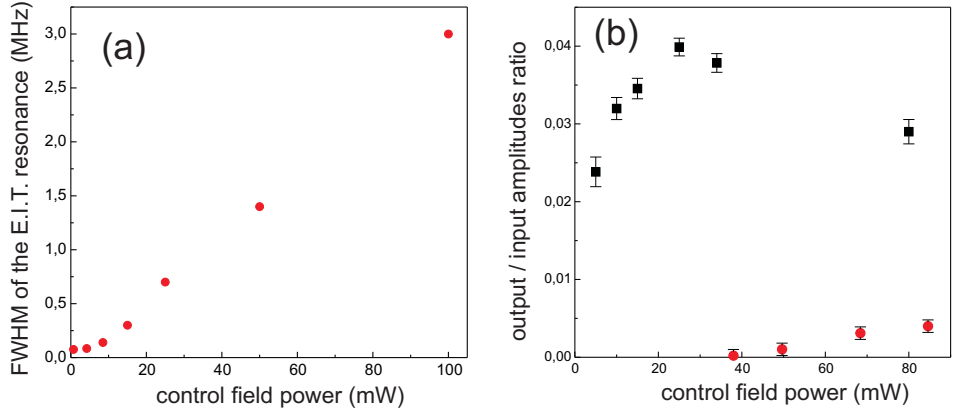}
\caption{ (a) Full width at half maximum of the EIT transparency
window (MHz) as a function of the control field power (mW). The
control and probe beams 1/$e^2$ radius is $7.2$~mm. (b) Ratio of
the output and input states amplitudes, as a function of the
control field power, for a single sideband of frequency 1.25 MHz
(squares) and a dual sideband modulation of frequency 400 kHz. T=
50~$^\mathrm{o}$~C . Pulse duration : 5 $\mu \mathrm{s}$. Storage
time : 15~$\mu \mathrm{s}$.} \label{width_power}
\end{figure*}

In this experiment we store a very weak coherent field that, as
explained before, can be considered as a single modulation sideband
of the control field. This  allows more flexibility than storing two
symmetrical sidebands in the same EIT window, especially for high
frequency components. We have measured the efficiency of the process
when the sideband frequency $\Omega$ is varied, as can be seen in
Fig.~\ref{st}. No significant variation with $\Omega$ is observed.
This comes from the fact that the frequency of the signal to be
stored can be matched to the position of the EIT window, without
changing its width. The EIT window width can be kept below 1 MHz, as
shown in Fig.~\ref{width_power}(a).

The two sidebands of a field can also be stored at the same time
if $2\Omega$ is smaller than the EIT linewidth. Figure
\ref{width_power}(a) shows the EIT linewidth in our case as a
function of the control field power. For a control field power
$P_{c}\sim 40$~mW, the full width at half maximum of the EIT
resonance is of the order of 1~MHz. We have measured the
efficiency of the storage process for a signal field made of two
sidebands at $\pm$400~kHz. The result is shown in
Fig.~\ref{width_power}(b). The efficiency is lower than in the
single sideband case, which can be attributed to the large value
of the time-bandwith product of the pulse to be stored.

\section{Noise characteristics}

\begin{figure*}[ht]
\centering
\includegraphics[width=0.8\textwidth]{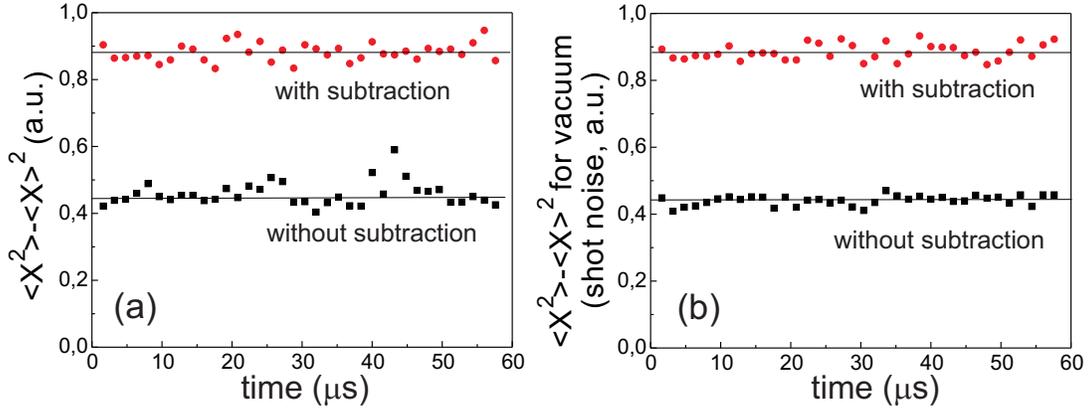}
\caption{ Variance of the amplitude quadrature for the signal (a)
and for the vacuum (b). The units are arbitrary and are the same in
(a) and (b) and for the mean values squared in Fig. \ref{signal}.}
\label{variances}
\end{figure*}

A critical feature for a quantum memory is the noise characteristics
of the outgoing signal. The  noise curves shown in Fig.
\ref{variances}(a) correspond to the mean values shown in Fig.
\ref{signal} and are obtained by calculating the variances from the
same data set. Because of a small leak of the control field into the
signal field channel, the raw data exhibit additional features due
to the transients of the control field. Although it has been
designed with a smooth shape, the latter contains Fourier components
around $\Omega/2\pi=1.2$ MHz. To get rid of this spurious effect, we
have used a subtraction procedure. The transients are measured
independently after each sequence with no signal field and the
corresponding data are subtracted point to point from the data taken
with a signal field. The signal curves shown in Fig. \ref{signal}
are obtained with this method. For the noise, this procedure is
equivalent to a 50/50 beamsplitter on the analyzed beam and adds one
unit of shot noise to the noise that would be measured without the
subtraction, yielding the upper curve in Fig. \ref{variances}(a) for
the variance of the amplitude quadrature. The variance of the phase
quadrature behaves in a similar way. The noise calculated from the
raw data, without subtraction, corresponding to the lower curve in
Fig. \ref{variances}(a) exhibits a small amount of additional
fluctuations when the control field is turned on for read-out. With
the subtraction procedure, these fluctuations are suppressed,
showing that they originate from a classical, reproducible spurious
effect.

The noise curves can be compared to the shot noise, which is
obtained independently from the same procedure with no control
field and no signal field in the input, without and with
subtraction, as shown in Fig.~\ref{variances}(a). The recorded
variances shown in Fig.~\ref{variances}(a) are found to be at the
same level as shot noise (Fig.~\ref{variances}(b)), which
indicates that the writing and reading processes add very little
noise. From these measurements, excess noise can be evaluated to
be zero with an uncertainty of less than 2$\%$. The noise curves
of Fig.~\ref{variances}(a) correspond to moderate values of the
control field power (10~mW). When the control field power is set
to higher values, excess noise appears. Excess noise has been
studied by other authors \cite{Lam,Hetet}. In our case, it
originates from fluorescence and coherent emission due to the
control beam \cite{Lvovsky} and from spurious fluctuations from
the turn on of the control field leaking into the signal channel,
that cannot be eliminated with the subtraction procedure.

Following Ref.~\cite{Hetet}, we can evaluate the quantum
performance of our storage device using criteria derived from the
T-V characterization.  This method was proposed in
Ref.~\cite{Roch} to characterize quantum non demolition
measurements or teleportation, and allows to obtain a state
independent quantum benchmark. The conditional variance product
$V$ of the signal field quadratures before and after storage is
the geometrical mean value of the input-output conditional
variances $V=\sqrt{V_{X}V_{Y}}$ of the two quadratures, with

\begin{eqnarray}
V_{X}=V_{X}^{out}-\frac{|\langle\hat{X}^{in}\hat{X}^{out}\rangle|^{2}}{V_{X}^{in}}\\
V_{Y}=V_{Y}^{out}-\frac{|\langle\hat{Y}^{in}\hat{Y}^{out}\rangle|^{2}}{V_{Y}^{in}}
 \label{variance}
\end{eqnarray}
where $V_{X}^{in/out}$ ($V_{Y}^{in/out}$) is the variance of the
normalized input/output field quadratures denoted
$\hat{X}^{in/out}$ ($\hat{Y}^{in/out}$). The total signal transfer
coefficient $T$ is the sum of the transfer coefficients for the
two quadratures

\begin{equation} T=T_{X}+T_{Y}
 \label{T}
\end{equation}
with
\begin{equation}
T_{X}=\frac{\mathcal{R}_{X}^{out}}{\mathcal{R}_{X}^{in}}\hspace{1cm}
T_{Y}=\frac{\mathcal{R}_{Y}^{out}}{\mathcal{R}_{Y}^{in}}
 \label{TX}
\end{equation}
where $\mathcal{R}_{X}^{in/out}$ ($\mathcal{R}_{Y}^{in/out}$) is
the signal to noise ratio of input/output field for the $X$ ($Y$)
quadrature

\begin{equation}\label{Rx}
\mathcal{R}_{X}^{in/out}=
\frac{4(\alpha_{X}^{in/out})^{2}}{V_{X}^{in/out}}\hspace{1cm}
\mathcal{R}_{Y}^{in/out}=
\frac{4(\alpha_{Y}^{in/out})^{2}}{V_{Y}^{in/out}}
\end{equation}
where $\alpha_{X}^{in/out}$ ($\alpha_{Y}^{in/out}$) is the mean
amplitude of the field
 $\hat{X}^{in/out}$ ($\hat{Y}^{in/out}$) quadrature .

\begin{figure}[htpb!]
\centering
\includegraphics[width=0.9\columnwidth]{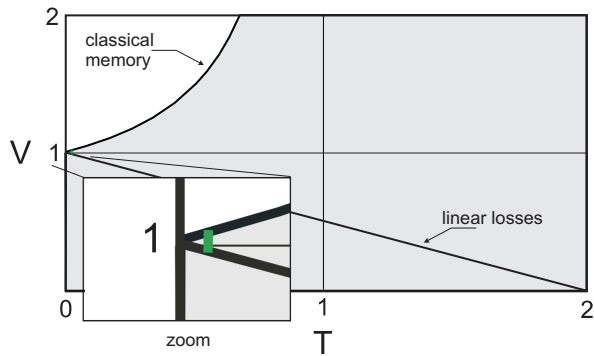}
\caption{T-V diagram. Plot of the conditional variance product $V$
versus total signal transfer coefficient $T$. The upper curve
shows the optimal performance of a classical memory. The linear
loss curve corresponds to an unperfect memory register that
introduces losses but no excess noise.} \label{TV}
\end{figure}

The experimental results shown in Fig.~\ref{signal} and
\ref{variances}, with a storage amplitude efficiency of 10$\%$,
correspond to $T=0.02$ and $V=0.99$, if one considers that the
noise is equal to shot noise, and $V=1.01$ if the noise is 2$\%$
higher than shot noise. This case is represented by the bar
represented in Fig. \ref{TV}.  A classical memory with the same
$T$, represented by the upper curve in Fig. \ref{TV}, yields
$V=1.01$. So, in this case, the performances of our system are
within the limit of the quantum domain. When the storage amplitude
efficiency is 21$\%$, which is obtained with a smaller storage
time and a higher control field, $T=0.08$, while the value of the
conditional variance of a system with such a $T$ without excess
noise is $V=0.96$ and the conditional variance for a classical
memory is $V=1.04$. The excess noise in our experiment in that
case is over 10$\%$, which corresponds to a performance well into
the classical domain. With the experimental parameters used here,
models \cite{Sorensen} predict much higher efficiencies, of the
order of 35$\%$ and no excess noise. The discrepancy can be
attributed to the fact that the D2 line of Cesium is far from a
simple $\Lambda$ scheme, and to decoherence effects in the lower
levels.

\section{Conclusion}

In this paper, we have studied the storage and retrieval of the
two non-commuting quadratures of a small coherent state in an
atomic medium with very small excess noise. This coherent state,
made of a single sideband of the control field, has been stored in
the Zeeman coherence of the atoms. The optimal response frequency
of the medium for storage can thus be adapted to the frequency to
be stored by changing the magnetic field, keeping the EIT window
rather narrow. Comparison with the storage of a modulation made of
two symmetrical sidebands shows the latter is hampered by the
finite bandwidth of the EIT window. Storing the two sidebands in
two separate windows is thus a promising method for quantum memory
with a widely adjustable frequency. For instance, in order to
store squeezed fields at a given frequency $\Omega$, one has to
store the two sidebands at $+\Omega$ and $-\Omega$. If $\Omega$ is
large, this can be done in two separate ensembles, after
separating the two sidebands using a Mach-Zehnder interferometer
\cite{Leuchs} or a cavity \cite{Ralph}. After read-out, the two
sidebands can be recombined. This procedure should lead to much
higher efficiency than increasing the EIT window.

\section{Acknowledgements}
This work was supported by the E. U. grants COVAQIAL and COMPAS,
by the French ANR contract IRCOQ and by the Ile-de-France
programme IFRAF. J.O. acknowledges support from the DGA and Dr. B.
Desruelle.


\begin{thebibliography}{10}

\bibitem{zoller} P. Zoller \textit{et al.},  Eur. Phys. J. D \textbf{36},
203-228 (2005).

\bibitem{kimble} H.J. Kimble, Nature \textbf{453},
1023-1030 (2008).

\bibitem{Polzik1} A.E. Kozhekin, K. M{\o}lmer, E.S. Polzik, Phys. Rev. A \textbf{62}, 033809
(2000).



\bibitem{DLCZ} L.M. Duan, M.D. Lukin, J.I. Cirac and P. Zoller, Nature (London) \textbf{414}, 413
(2001).



\bibitem{Dantan1} A. Dantan, M. Pinard, V. Josse, S. Nayak, P.R. Berman, Phys.
Rev. A \textbf{67}, 045801 (2003).



\bibitem{Lukin1} M.D. Lukin, S.F. Yelin and M. Fleischauer, Phys. Rev. Lett. \textbf{84}, 4232
(2000).


\bibitem{Lukin2} M. Fleischauer and M.D. Lukin, Phys. Rev. Lett. \textbf{84}, 5094
(2000).


\bibitem{Dantan2} A. Dantan and M. Pinard, Phys. Rev. A \textbf{69}, 043810
(2004).

\bibitem{Polzik2} K. Hammerer, K. M{\o}lmer, E.S. Polzik, J.I. Cirac, Phys.Rev. A \textbf{70},
044304 (2004).

\bibitem{Moiseev} A. Moiseev and S. Kroll, Phys. Rev. Lett. \textbf{87}, 173601 (2001)

\bibitem{Kraus} B. Kraus, W. Tittel, N. Gisin , M. Nilsson, S. Kroll, J.I.
Cirac, Phys. Rev. A \textbf{73} 020302(R) (2006)

 \bibitem{Sangouard} N. Sangouard, C. Simon, M. Afzelius, N. Gisin, Phys. Rev. A \textbf{75}, 032327
 (2007); H. de Riedmatten, M. Afzelius, M.U. Staudt, C. Simon, and N. Gisin,
 Nature \textbf{456}, 773 (2008)

\bibitem{james} C. W. Chou, H. de Riedmatten, D. Felinto, S.V. Polyakov, S.J. van Enk, H.J. Kimble, Nature
\textbf{438}, 828 (2005)

 \bibitem{julien}J. Laurat et al., New J. Phys.  \textbf{9},
207 (2007).

 \bibitem{james2}C.W. Chou et al., Science \textbf{316},
1316 (2007).

\bibitem{Pan} Y.-A. Chen et al.,
Nature Phys. \textbf{4}, 103 (2008)


\bibitem{Vuletic} J.K. Thompson, J. Simon, H. Loh, V. Vuletic, Science \textbf{313}, 74 (2006)

\bibitem{Lukin3} M.D. Eisaman, A. Andre, F. Massou, M. Fleischhauer, A.S. Zibrov, M.D. Lukin, Nature \textbf{438}, 837
(2005).

\bibitem{Kuzmich} T. Chaneliere, D.N. Matsukevich, S.D. Jenkins, S.Y. Lan, T.A.B. Kennedy, A. Kuzmich, Nature \textbf{438}, 833
(2005).

\bibitem{Choi} K.S. Choi, J. Laurat, H. Deng, H.J. Kimble,  Nature \textbf{452}, 67 (2008)

\bibitem{Zhao} B. Zhao,  Y.-A. Chen, X.-H. Bao, T. Strassel, C.-S.
Chuu, X.-M. Jin,  J. Schmiedmayer,  Z.-S. Yuan,  S. Chen, and
J.-W. Pan, Nature Physics \textbf{5}, 95 (2009)

\bibitem{Grangier} F. Grosshans, G. Van Assche, J. Wenger, R. Brouri, N. J. Cerf, and
P. Grangier, Nature \textbf{421}, 238 (2003)


\bibitem{Polzik3} B. Julsgaard, J. Sherson, J. Fiurasek, J.I. Cirac, E.S. Polzik,
Nature \textbf{432}, 482 (2004).


\bibitem{Kozuma} K. Honda, D. Akamatsu, M. Arikawa, Y. Yokoi, K.
Akiba, S. Nagatsuka, T. Tanimura, A. Furusawa, M. Kozuma, Phys. Rev.
Lett. \textbf{100}, 093601 (2008).


\bibitem{Lvovsky} J. Appel, E. Figueroa, D. Korystov, A.I.
Lvovsky, Phys. Rev. Lett. \textbf{100}, 093602 (2008).

\bibitem{Cviklinski} J. Cviklinski, J. Ortalo, J. Laurat, A. Bramati, M. Pinard, E. Giacobino,
 Phys. Rev. Lett. \textbf{101}, 133601 (2008)


\bibitem{Harris} S. E. Harris, J. E. Field, and A. Imamoglu, Phys. Rev. Lett. \textbf{64}, 1107 (1990);
 S.E. Harris Phys. Today \textbf{50}, 36 (1997).


\bibitem{Hau1} L.V. Hau, S. E. Harris, Z. Dutton and C.H. Behroozi, Phys. Rev. Lett. \textbf{82}, 4611
(1999).


\bibitem {Hau2} C. Liu, Z. Dutton, C.H. Behroozi and L.V. Hau, Nature
(London) \textbf{409}, 490 (2001).


\bibitem {Phillips} D.F. Phillips, A. Fleischhauer, A. Mair, R.L.
Walsworth and M.D. Lukin, Phys. Rev. Lett. \textbf{86}, 783 (2001).


\bibitem {Dantan3} A. Dantan, J. Cviklinski, M. Pinard and Ph. Grangier, Phys. Rev. A, \textbf{73}, 032338
(2006).

\bibitem{Sorensen} A.V. Gorshkov, A. Andr\'{e}, M.D. Lukin, A.S. S{\o}rensen, Phys. Rev. A \textbf{76},
033805 (2007).

\bibitem {Cusack} B.J. Cusack, B.S. Sheard, D.A. Shaddock, M.B. Gray, P. K. Lam, S.E. Whitcomb, Appl. Opt. \textbf{43}
5079 (2004).

\bibitem{coherence} A. Mair, J. Hager, D. F. Phillips, R. L. Walsworth and M. D.
Lukin, Phys. Rev. A, \textbf{65}, 031802(R), (2002).

\bibitem {Lam} M.T.L. Hsu, G. H\'{e}tet, O. Gl\"{o}ckl, J.J. Longdell, B.C. Buchler, H.A.
Bachor, P.K. Lam, Phys. Rev. Lett. \textbf{97}, 183601 (2006).


\bibitem {Caves} C.M. Caves, Phys. Rev. D\textbf{26}, 1817, (1982); J.
Gea-Banacloche and G. Leuchs, J. Mod. Opt. \textbf{34}, 793 (1987)

\bibitem {Leuchs} O. Gl\"{o}ckl, U. L. Andersen, S. Lorenz, Ch. Silberhorn, N. Korolkova, and G.
Leuchs, Opt. Lett., \textbf{29}, 1936 (2004).

\bibitem {Ralph} E. H. Huntington and T. C. Ralph, J. Opt. B: Quantum Semiclass. Opt. \textbf{4} 123–128 (2002).

\bibitem {Hetet} G. H\'{e}tet, A. Peng, M. T. Johnsson, J. J. Hope, and P. K. Lam \textit{et al.}, Phys. Rev. A \textbf{77} 012323
(2008).

\bibitem {Roch} J.-F. Roch, K. Vigneron, Ph. Grelu, A. Sinatra, J.-Ph. Poizat, and
Ph. Grangier, Phys. Rev. Lett. \textbf{78} 634 (1997); T.C. Ralph
and P.K. Lam, Phys. Rev. Lett. \textbf{81} 5668 (1998).


\end{thebibliography}
\end{document}